\documentclass[5p]{elsarticle}

\usepackage{lineno}
\usepackage{hyperref}
\usepackage{amsmath}
\let\oldcdot\cdot
\usepackage{breqn}
\let\cdot\oldcdot
\usepackage{multirow}
\usepackage{siunitx}
\usepackage{collcell}
\usepackage{booktabs}
\usepackage{textgreek}
\usepackage{textcomp}

\modulolinenumbers[5]

\journal{Nuclear Instruments and Methods in Physics Research Section A}

\newcommand\Tstrut{\rule{0pt}{2.6ex}}
\newcommand\Bstrut{\rule[-0.9ex]{0pt}{0pt}}

\begin{document}

\begin{frontmatter}

\title{Radon Daughter Plate-out Measurements at SNOLAB {\tiny }for Polyethylene and Copper}

\author[smuaddress]{Matthew Stein}
\ead{mstein@smu.edu}

\author[fnaladdress]{Dan Bauer}

\author[pnnladdress]{Ray Bunker}

\author[smuaddress]{Rob Calkins}

\author[smuaddress]{Jodi Cooley}

\author[pnnladdress]{Ben Loer}

\author[snolabaddress]{Silvia Scorza}

\address[smuaddress]{Department of Physics, Southern Methodist University, Dallas, TX 75205, USA}
\address[fnaladdress]{Fermi National Accelerator Laboratory, Batavia, IL 60510, USA}
\address[pnnladdress]{Pacific Northwest National Laboratory, Richland, WA 99352, USA}
\address[snolabaddress]{SNOLAB, Lively, Ontario P3Y 1N2, Canada}

\begin{abstract}
\begin{sloppypar}
Polyethylene and copper samples were exposed to the underground air at SNOLAB for approximately three months
while several environmental factors were monitored. Predictions of the radon-daughter plate-out rate are compared to the resulting surface activities, obtained from high-sensitivity measurements of alpha emissivity using the XIA UltraLo-1800 spectrometer at Southern Methodist University.  From these measurements, we determine an average $^{210}$Pb plate-out rate of \mbox{249 and 423~atoms/day/cm$^{2}$} for polyethylene and copper, respectively, when exposed to radon activity concentration of 135 Bq/m$^{3}$ at SNOLAB. A time-dependent model of alpha activity is discussed for these materials placed in similar environmental conditions.
\end{sloppypar}

\end{abstract}


\end{frontmatter}


\section{Introduction}
\begin{sloppypar}
Many low-background experiments
are placed deep underground to shield from cosmic rays.
Extreme care is taken to account for and to avoid accumulation of radiocontaminants on material surfaces, which is made all the more challenging by the high radon activity typically present in underground laboratories.
Radon daughters in air can plate out onto and implant within experiment materials, and these daughters can give rise to neutron and gamma-ray backgrounds from ($\alpha$,n) and Bremsstrahlung interactions, respectively. Useful metrics for this implantation process are the plate-out rate (implanted atoms/area/time) or plate-out height (height under which it is assumed 100\% of all radon daughters will plate onto the surface below).
\end{sloppypar}
\begin{sloppypar}
After a series of short ($<1$~hr) decays, $^{210}$Pb ($t_{1/2}=22.3$ yr) comprises the majority of remaining contaminants. This isotope decays via $\beta$ emission to $^{210}$Bi, which subsequently $\beta$ decays ($t_{1/2}=5$~d) to $^{210}$Po. This study focuses on the $5.3$ MeV alphas from $^{210}$Po decays ($t_{1/2}=138$~d). These alphas can interact with $^{13}$C nuclei in polyethylene (commonly used as a neutron shield) and generate neutrons through ($\alpha$,n) reactions. For dark matter direct detection experiments, neutrons are a challenging background because they deposit energy in a way that can mimic the signals expected from dark matter interactions.  
As a result, it is important to understand how contamination from radon and its progeny can lead to neutron backgrounds and how they evolve over time.
\end{sloppypar}
\section{Estimating Backgrounds in Polyethylene}


Several forms of polyethylene, (C$_2$H$_4$)$_n$, are commercially available. This study focuses on high-density poly-ethylene (HDPE) which has density of $0.941\textrm{--}0.965$ g/cm$^3$. With a natural abundance of $1.07(8)\%$ \cite{nist}, $^{13}$C accounts for $0.36\%$ of all atoms in HDPE. 

\begin{figure}
	\caption{SOURCES-4C neutron spectrum from 1 Bq of $^{210}$Pb contamination in each gram of polyethylene (assuming $^{210}$Po is in secular equilibrium).}
	\includegraphics[width=\linewidth]{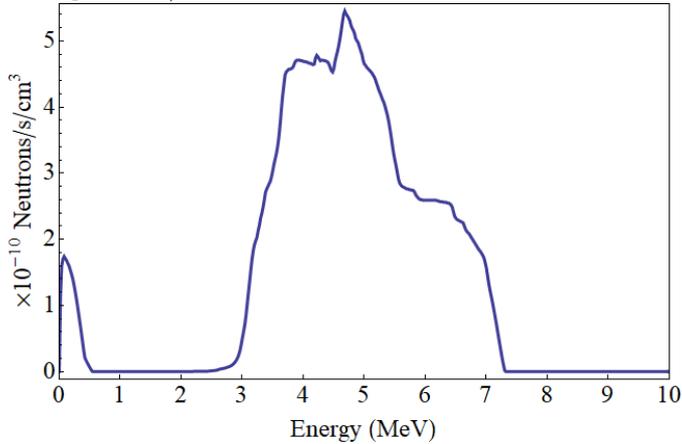}
	\label{im_PoN_1ppbPb}
\end{figure}

A modified version of SOURCES\nobreakdash-4C \cite{s4c,modsources}
was used to model ($\alpha$,n) reactions in HDPE,
resulting in an expectation of 7.3$\times$10$^{-8}$~n/s/cm$^3$ for 1 Bq/g of $^{210}$Pb activity (assuming secular equilibrium) in the bulk of the polyethylene (\textit{cf.}\ Fig. \ref{im_PoN_1ppbPb}).  
Polyethylene shielding exposed to a high-radon environment such as SNOLAB would quickly become contaminated with residual $^{210}$Pb.  Though the $^{210}$Pb would be implanted near the surface, and some alphas from the $^{210}$Po decays would be emitted away from the bulk, there is still the possibility these alphas could interact with $^{13}$C on an exiting trajectory.  For the purposes of a conservative estimate, any alpha activity on or near the surface is considered as having the potential to create neutron backgrounds.


\section{Experimental Setup \& Environment}

\subsection{Cavern Environment}

\begin{figure}
	\caption{Radon activity per m$^3$ at SNOLAB during the exposure period as measured by a RadStar RS300.  The average measured value of 135\,Bq/m$^3$ (green line) and associated 1$\sigma$ standard deviation (shaded band) are also shown.}
	\includegraphics[width=\linewidth]{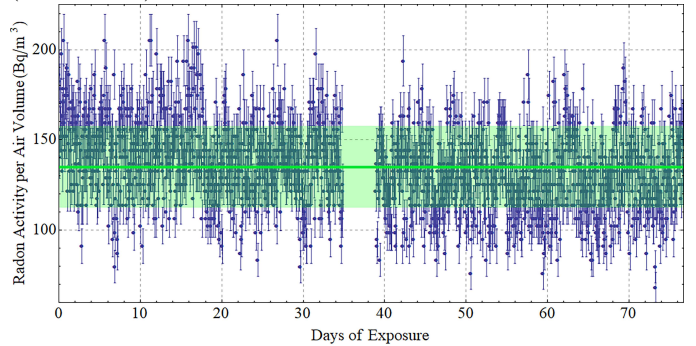}
	\label{im_Radon}
\end{figure}

The experimental site is located at SNOLAB, a Class 2000 clean room laboratory 6,800 feet below the surface in Lively, Ontario, Canada.
The 
setup was
located in Room 127 in an area referred to as the Ladder Labs (see Fig.~\ref{im_map4}). 
During the exposure at SNOLAB, environmental factors were continuously monitored including radon, temperature, relative humidity, and counts of dust particles $\ge$0.3~$\mu$m.  The instruments used to record these data were located on a table immediately adjacent to the samples at Site 1 (see Table \ref{table:EnvData} and Figs. \ref{im_Radon}, \ref{im_map4}).  We monitored these values such that we could either rule out or include possible effects from fluctuating environmental factors. Radon levels in the laboratory area are known to seasonally vary from ${\sim}125$ to $135$ Bq/m$^3$ whereas the the level at the surface is around 6 Bq/m$^{3}$ \cite{SNOLABHandbook}.
\begin{table}
	\centering
	\begin{tabular}{lcc}\hline
		\Tstrut Data & Average & $\sigma$\\
		\hline\Tstrut
		Particles $\geq 0.3 \mu$m (pp. ft$^3$) & $238$ & $679$\\[1.9pt]
		Radon (Bq/m$^3$) & $135$  & $23$\\[1.9pt]
		Temperature (K) & $293.3$ & $0.4$\\[1.9pt]
		Humidity (\%) & $57.9$ & $1.6$\\\hline
	\end{tabular}
	\caption{Average environmental values of the experiment location within \mbox{SNOLAB}, with one standard deviation calculated from the population of data points. Dust particles were monitored with a ParticleScan CR, radon activity with a RadStar RS300, and temperature and humidity with a \mbox{Lascar} EL-USB-2-LCD+.
	The large particle-count standard deviation results from (a few) intermittent periods during which the level briefly exceeded 10$^4$ $\geq$0.3~$\mu$m particles per ft$^3$. This was observed in the vicinity of the experimental setup and is likely due to installation of wiring in a nearby area and some pressure testing of copper lines.  However, throughout the exposure period, the SNOLAB (lab wide) particle monitors (which measure particles $>$~0.5$\mu$m) recorded levels consistent with a Class 2000 environment or better.\label{table:EnvData}}
\end{table}

\subsection{Panels \& Placement}

\begin{figure}
	\caption{Map of the four exposure sites in the Ladder Labs at SNOLAB 
		(\textit{cf}. Table \ref{table:locations}).
	}
	\includegraphics[width=\linewidth]{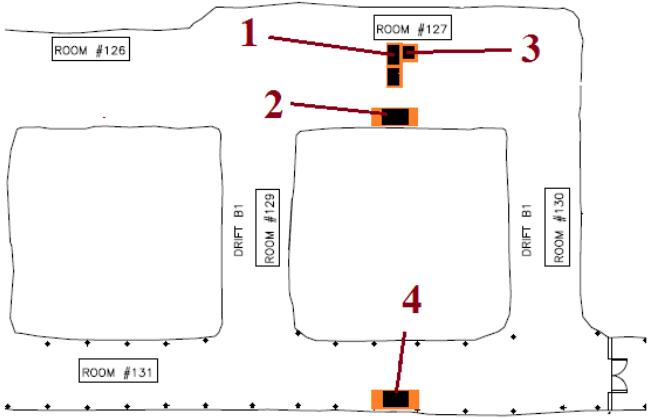}
	\label{im_map4}
\end{figure}

A total of ten HDPE panels were used, all cut from the same 122~cm$\times$244~cm sheet (purchased from Johnston Industrial Plastics, Ontario, Canada), each panel of dimensions \mbox{30.5~cm$\times$30.5~cm$\times0.5$~cm}.  We chose this sample size to optimize the sensitivity of the UltraLo-1800 spectrometer that was used to perform the pre- and post-exposure surface assays in this study.
The panels were set in pairs at four different locations in SNOLAB with varying height and room position to test for variations in plate-out from position and proximity to nearby walls
(see Table \ref{table:locations} and Fig. \ref{im_map4}).  Each pair was set immediately adjacent to one another with each panel laid flat.

Four copper panels were also placed at Site 1, each of dimension 15.25~cm$\times$30.5~cm$\times$0.64~cm.  Every panel was placed on a non-conducting surface for the duration of the exposure.
During shipment to and from SNOLAB, all panels were sealed inside two nitrogen-flushed static dissipative nylon bags with an outer polyethylene bag.  The polyethylene bag was used as a general protection around the inner bags while the nylon bags were chosen for their low radon permeability \cite{Borexino2004}.

For the trip to SNOLAB, the panels were laid face-to-face with no air gaps.  For the return trip, the panels were packed in pairs (one pair from each site, upward-facing sides pointed inward) with a small air gap between panels to best maintain the integrity of the surfaces.  The bags were once again nitrogen back-filled to limit any plate-out that might occur during shipment.  

\begin{table}
	\centering
	\begin{tabular}{cccc}\hline
		\Tstrut Site & Room & Nearest & Height\\
		Number & Number & Wall (m) & (m) \\\hline
		\Tstrut $1$ & 127 & $3.63$ & $0.94$\\[1.9pt]
		$2$ & 127 & $0.38$ & $0.94$\\[1.9pt]
		$3$ & 127  & $3.63$ & $2.01$\\[1.9pt]
		$4$ & 131 & $0.38$ & $0.94$\Bstrut\\\hline
	\end{tabular}
	\caption{Position information for each exposure location used.  Height is measured as the distance from the floor to the surface of the panels.  Two polyethylene samples were placed at each location and four copper samples were placed at Site 1.  The variety of locations was motivated to test for variations in plate-out height due to position and proximity to nearby walls.\label{table:locations}}
\end{table}

\section{Analysis \& Results}

\subsection{Pure $^{210}$Po Model}

Because the expected alphas come from the short-lived daughter ($^{210}$Po) of a long-lived parent ($^{210}$Pb), a model for the number of $^{210}$Po atoms over time is built from the Bateman equation \cite{bateman}:
\begin{equation}
N_{Po}(t)=N_{Pb}(0)\frac{\lambda_{Pb}}{\lambda_{Po}-\lambda_{Pb}}\left(e^{-\lambda_{Pb}t}-e^{-\lambda_{Po}t}\right)
\label{eq:Bateman}
\end{equation}
Due to the comparatively short half-life of $^{210}$Bi ($\sim5$~days), we neglect it in setting up Equation \ref{eq:Bateman}.  To account for a particular number of $^{210}$Pb atoms ($N_i$) added at a specific non-zero time $t_i$, Equation \ref{eq:Bateman} can be altered as follows:
\begin{equation}
\begin{aligned}
N_{Po,i}(t,t_i,N_i)={}&N_i\frac{\lambda_{Pb}}{\lambda_{Po}-\lambda_{Pb}}\left(e^{-\lambda_{Pb}(t-t_i)}-e^{-\lambda_{Po}(t-t_i)}\right)\\
&\times\Theta\left(t-t_i\right)
\label{eq:ModBateman}
\end{aligned}
\end{equation}
Note that because of the relatively stable radon activity at SNOLAB during this exposure (\textit{cf.}\ Fig. 2), the radon-daughter plate-out rate onto the sample surfaces is approximately constant with time; thus we assume that $^{210}$Pb atoms are being added at a constant rate $R_{{Pb}}$.
The total number of $^{210}$Po atoms is then a sum of Equation (\ref{eq:ModBateman}) over an exposure period ($t_{exp}$) in (preferably small) step sizes ($b\equiv$ time-step size). 
 
\begin{figure}
	\caption{Four individual forms of Equation (\ref{eq:ModBateman}) (blue, red, yellow, and green curves) summed up (black dashed curve), showing exponential growth during exposure, and then later coming into secular equilibrium. Inset more clearly shows exponential growth during exposure period which in this example ends at a time of 150, and the step size is 50.}
	\includegraphics[width=\linewidth]{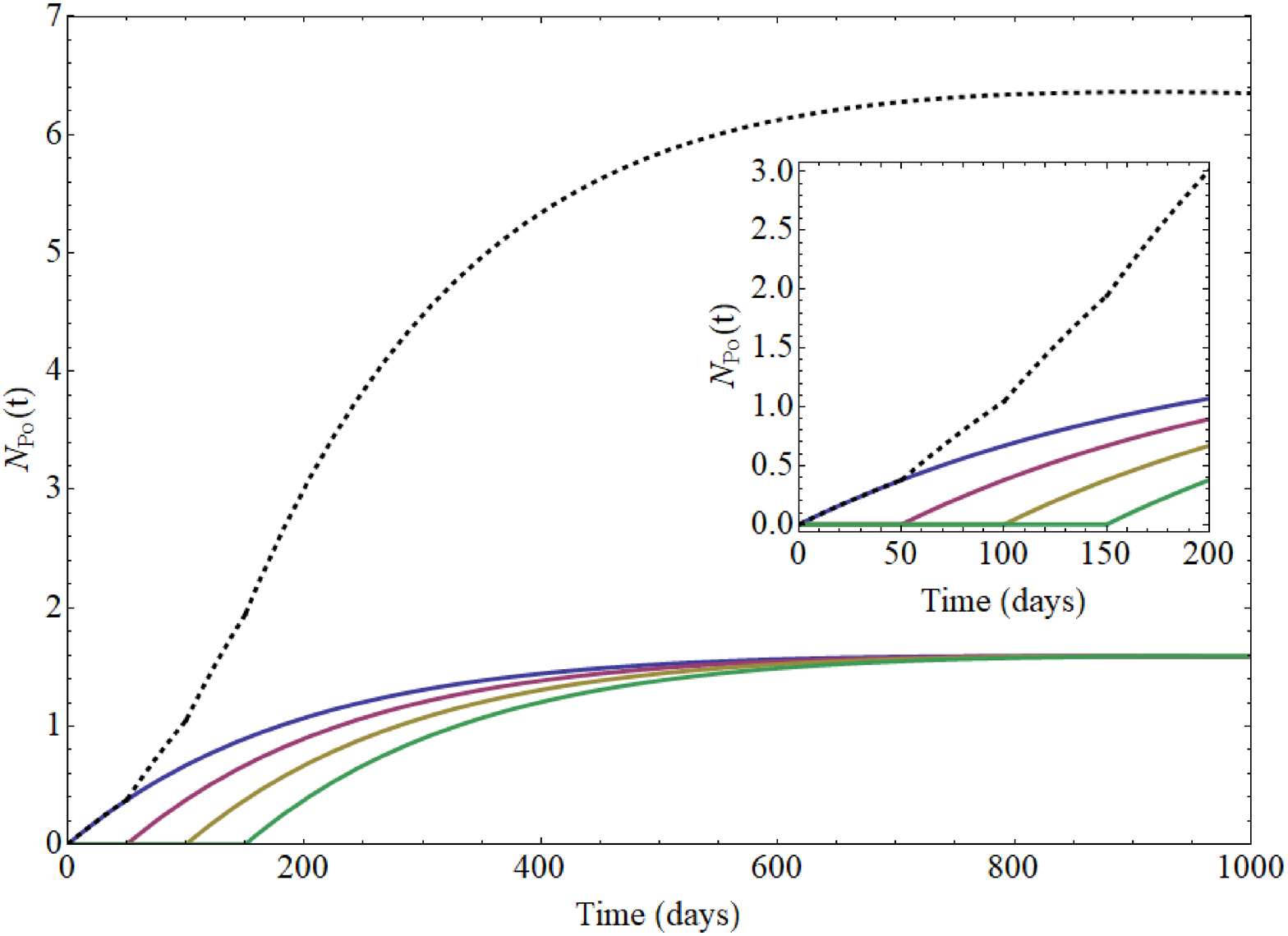}
	\label{im_4batemans_combined}
\end{figure}

As an example, consider a 150~day exposure in an environment where $R_{Pb}=100$~atoms/cm$^2$/day. Rather than assuming all the atoms plate out at once, the exposure can be broken down into four depositions separated by 50~days (see Fig. \ref{im_4batemans_combined}), each deposition taking the form of Equation~(\ref{eq:ModBateman}). Smaller gaps between depositions will result in a more accurate total value, both during and after exposure.  The total activity of $^{210}$Po is then:
\begin{equation}
\begin{aligned}
A_{Po}(t,t_{exp})={}&\lambda_{Po}\left[\sum_{i=1}^{t_{exp}/b}N_{Po,i}(t,i\times b,R_{Pb}\times b)\right]
\label{eq:ActBateman}
\end{aligned}
\end{equation}

Making the substitution of $n\equiv\frac{t_{exp}}{b}$ and taking the limit of Equation (\ref{eq:ActBateman}) as $n\rightarrow\infty$, the following closed-form expression describes the activity over time:

\begin{dmath}
{A_{Po}(t,t_{exp})=}\frac{R_{Pb}}{\lambda_{Pb}-\lambda_{Po}}\\\cdot\left[{\lambda_{Pb}\left(1-e^{-\lambda_{Po}t}\right)+\lambda_{Po}\left(e^{-\lambda_{Pb}t}-1\right)}+\Theta\left(t-t_{exp}\right)\\\cdot\left(\lambda_{Pb}\left\{e^{-\lambda_{Po}\left(t-t_{exp}\right)}-1\right\}+\lambda_{Po}\left\{1-e^{-\lambda_{Pb}\left(t-t_{exp}\right)}\right\}\right)\right]\raisetag{-5em}
\label{eq:ActBatemanClosed}
\end{dmath}



Knowing the exposure time $t_{exp}$ and measurement time $t_{m}$, one can measure $A_{Po}(t_{m},t_{exp})$ and solve Equation (\ref{eq:ActBatemanClosed}) for $R_{Pb}$. 

\subsection{Including Long-Lived Activity}

\begin{figure}
	\caption{The model of total activity (solid red curve, Equation \ref{eq:ActTotal}) from $^{210}$Po (dashed yellow curve) and dust (dashed green curve).  The vertical line indicates the end of the 83 day exposure of our HDPE and copper samples to the SNOLAB environment. The case of no contribution from dust is also shown (dot-dashed blue curve, Equation \ref{eq:ActBateman}).}
	\includegraphics[width=\linewidth]{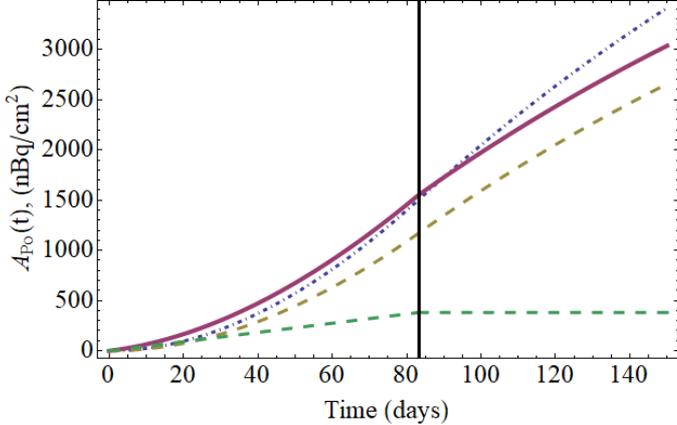}
	\label{im_model_plus_const}
\end{figure}

A model of total activity should consider the possibility that long-lived activity from dust may be present on the sample surfaces.
This study is primarily concerned with the dust's U and Th content. Because U and Th are so long-lived, their decay chains (assuming secular equilibrium) would contribute an approximately constant rate of radioactivity over the timescale of any experiment for a given amount of dust\footnote{Although there is $^{210}$Po and $^{210}$Pb activity in the dust, we assume that it is in secular equilibrium with the rest of the $^{238}$U decay chain.  As such, a simpler global treatment of the time dependence is possible that does not need to explicitly track the $^{210}$Po and $^{210}$Pb half-lives.}. 
Consequently, if dust settles at a constant rate per unit time $S_{{dust}}$, then the total activity from dust ($A_{{dust}}$) should accumulate linearly during exposure to the SNOLAB environment and then remain constant once the exposure concludes: 

\begin{equation}
A_{{dust}}(t,t_{exp})=\begin{cases}
S_{{dust}}t&t < {t_{exp}}\\
S_{{dust}}t_{exp}&t\geq{t_{exp}}\\
\end{cases}
\label{eq:dust}
\end{equation}
The total activity from all sources --- $^{210}$Po and dust (see Fig. \ref{im_model_plus_const}) --- is then
\begin{equation}
\begin{aligned}
A_{T}(t,t_{exp})={}&A_{Po}(t,t_{exp})+A_{{dust}}(t,t_{exp})
\label{eq:ActTotal}
\end{aligned}
\end{equation}
The values $R_{Pb}$ and $S_{{dust}}$ can be separated from the other parts of $A_{Po}(t,t_{exp})$ and $A_{{dust}}(t,t_{exp})$ respectively, 
yielding time-dependent functions that also depend on the exposure time:
\begin{equation}
A_{T}(t,t_{exp})=R_{Pb}f(t,t_{exp})+S_{{dust}}g(t,t_{exp})\nonumber
\end{equation}

With two measurements of activity spaced adequately apart ($t=t_{1}$,$t=t_2$), and with a known exposure time, one can then solve a linear system of equations for $R_{Pb}$ and $S_{{dust}}$:

\begin{equation}
\begin{pmatrix}
A_{T}(t_1,t_{exp}) \\
A_{T}(t_2,t_{exp})
\end{pmatrix}
=
\begin{pmatrix}
f(t_1,t_{exp}) & g(t_1,t_{exp}) \\
f(t_2,t_{exp}) & g(t_2,t_{exp})
\end{pmatrix}
\begin{pmatrix}
R_{Pb} \\
S_{{dust}}
\end{pmatrix}
\label{eq:LinSystem}
\end{equation}

\subsection{Measurements}

\begin{figure}
	\caption{Efficiency-corrected surface alpha activity of one HDPE sample following an 83-day exposure underground at SNOLAB, measured 10 and 90~days after the end of the exposure (``Meas. 1'' and ``Meas. 2'' respectively). There is a clear $^{210}$Po peak centered at 5.3~MeV. The low-energy tail is more extensive than that expected from TRIM simulations, so this may correspond to energy losses from surface roughness.}
	\includegraphics[width=\linewidth]{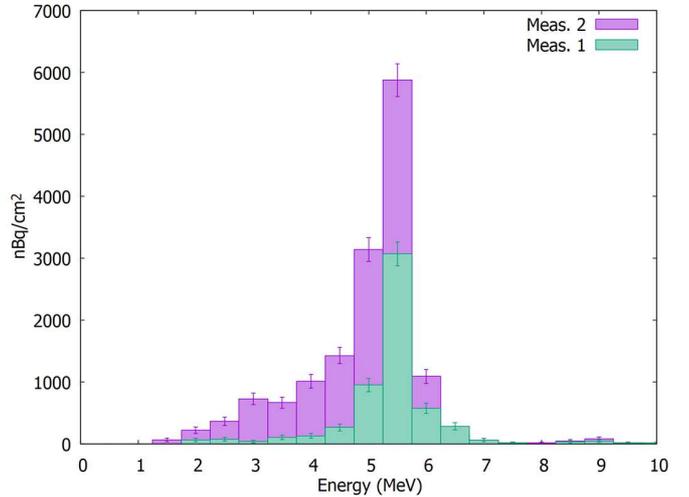}
	\label{im_poly3after}
\end{figure}

The simulation program TRIM \cite{SRIM} was used to simulate the implantation of radon daughters into polyethylene and copper.  The exiting energy of alphas from $^{210}$Po decays were also examined.  From this simulation, we expect $\sim98$\% of all exiting alphas to have energy within the 2.0--5.8\,MeV range after taking the UltraLo-1800 resolution into account ($\textless9$\% FWHM, Fig. \ref{im_poly3after}).  In this study, calculations of total alpha activity from each panel are made by integrating over this energy range.

Additionally, we investigated the feasibility of assaying HDPE with the UltraLo-1800 by measuring an HDPE sample with a calibrated $^{230}$Th source placed on top of the sample. The observed rate was consistent with the source's calibrated activity, suggesting that any outgassing from the sample was small enough to not significantly affect the detector response. XIA notes that measurement of non-conductive samples may affect the detector response due to distortion of the spectrometer's drift field, and indeed  we observed a slight degradation of the spectral resolution and a small bias toward lower energies; however, our choice of a relatively large region-of-interest minimizes any corresponding systematic uncertainty for the results presented here.

\subsubsection{Pre-exposure Assays}

After cleaning all copper and polyethylene samples with Radiacwash\texttrademark, deionized water, and isopropyl alcohol, three polyethylene panels (all from the same stock) and all four copper panels were assayed with the UltraLo-1800 to measure baseline activity.  Emissivity in the 2--10\,MeV and 2.0--5.8\,MeV ranges
was examined to determine the samples' surface activities prior to exposure underground at SNOLAB.  The results are summarized in Table \ref{table:PreExp} and the activities are sufficiently low to ensure that any surface contamination accumulated at SNOLAB will be clearly identifiable.

\begin{table}
	\centering
	\begin{tabular}{
		l
		S[table-format=3.1(3)]
		S[table-format=3.1(3)]
		}
		\hline
		\Tstrut Material & \multicolumn{2}{c}{Pre-exposure Activity}{}\\
		& \multicolumn{2}{c}{(nBq/cm$^2$)}{}\\
		\cmidrule{2-3} & {2--10\,MeV} & {2.0--5.8\,MeV} \\ \hline
		\Tstrut 
		HDPE & 187.5 \pm 25.6 & 97.2 \pm 18.4 \\[1.9pt]
		Copper & 524.9 \pm 71.1 &  393.7 \pm 61.6\Bstrut\\\hline
	\end{tabular}
	\caption{Initial alpha activity of polyethylene and copper samples after initial cleaning and prior to exposure at SNOLAB.\label{table:PreExp}}
\end{table}

\subsubsection{Control Samples}

Two polyethylene samples were used as control samples and not initially brought underground. They were instead left in a surface building at SNOLAB in their nylon bags.  At the very end of the exposure period, these control samples were brought underground, briefly removed from their bags, and packed with the other samples for the return trip to Southern Methodist University (SMU) where the alpha activity was measured.

After returning to SMU, the average alpha activity in
\makebox[\linewidth][s]{the 2--10(2.0--5.8)\,MeV range was determined to be}\\
$196.9\pm32.0$($90.4\pm21.7$) nBq/cm$^2$.  This was after $\sim$30~days of storage in an acrylic cabinet purged with low-radon liquid-nitrogen boil-off gas.
These levels are consistent with the average pre-exposure activities measured for the HDPE samples in Table \ref{table:PreExp}.
It is therefore concluded that no significant increase in activity was acquired from the shipping, transport, and storage of the samples.

\subsubsection{Exposed Samples}

Samples were rebagged and shipped back to SMU after an 83~day exposure at SNOLAB. During shipping, some of the HDPE sample pairs made contact in the very centers of the panels --- the intended air gap for each sandwiched pair was not perfectly maintained during the return trip.  Consequently, there may have been some cross-contamination between the active sample surfaces in each pair.  We believe that the effects of any such cross-contamination should be small.  However, because the pairs were co-located during the exposure, results can be pair-wise averaged to minimize the impact of this unintended cross-contamination.

Each sample was measured in the UltraLo-1800 spectrometer $\sim$10 days after the end of the exposure period, providing a first measure of the total surface alpha activity due to $^{210}$Po and dust. The samples were then rebagged and stored for $\sim$80 days in an acrylic cabinet purged with low-radon liquid-nitrogen boil-off gas. A second follow-up measurement was performed to obtain the time dependence of the surface activity, thus allowing Equation (\ref{eq:ActBatemanClosed}) to be solved for $R_{{Pb}}$ and $S_{{dust}}$; the estimated $^{210}$Pb and dust contamination rates are summarized in Table \ref{table:PostExp}. The measured spectra for one of the HDPE samples are shown in Fig. \ref{im_poly3after}, and all measured HDPE alpha rates are shown in Fig. \ref{im_AllPoints}.

\begin{table}
	\centering
	\small
	\begin{tabular}{l
			c
			S[table-format=3.1]@{\,\( \pm \)\,}S[table-format=2.1]
			S[table-format=2.1]@{\,\( \pm \)\,}S[table-format=2.1]
			S[table-format=2.1]@{\,\( \pm \)\,}S[table-format=1.1]}
		\hline
		\Tstrut 
		& & \multicolumn{2}{c}{$R_{Pb}$} & \multicolumn{2}{c}{$S_{{dust}}$} & \multicolumn{2}{c}{Plate-out} \\
		\mbox{Sample \#} & Site & \multicolumn{2}{c}{$\left(\frac{\textrm{atoms}}{\textrm{day}\cdot\textrm{cm}^2}\right)$} & \multicolumn{2}{c}{$\left(\frac{\textrm{nBq}}{\textrm{day}\cdot\textrm{cm}^2}\right)$} & \multicolumn{2}{c}{$\textrm{Height}\atop \textrm{(cm)}$} \\\hline
		\Tstrut 
		HDPE 1 & 1 & 257.0 & 26.2 & 19.6 &  8.1 & 22.0 & 2.2 \\[1.9pt]
		HDPE 2 & 1 & 334.0 & 31.7 & 15.8 & 10.6 & 28.6 & 2.7\\[1.9pt]
		HDPE 3 & 2 & 278.1 & 28.5 & 34.7 & 10.8 & 23.8 & 2.4\\[1.9pt]
		HDPE 4 & 2 &385.9 & 36.1 & 4.0 & 14.7 & 33.1 & 3.1\\[1.9pt]
		HDPE 5 & 3 & 155.6 & 33.6 & 69.4 & 12.5 & 13.3 & 2.9\\[1.9pt]
		HDPE 6 & 4 & 150.8 & 25.0 & 15.5 & 9.2 & 12.9 & 2.1\\[1.9pt]
		Copper 1 & 1 & 413.8 & 11.8 & 4.5 & 9.8 & 35.5 & 1.0\\[1.9pt]
		Copper 2 & 1 & 443.6 & 17.8 & 4.9 & 8.4 & 38.0 & 1.5\\\hline
		\Tstrut
		Avg. HDPE & & 248.6 & 12.0 & 24.9 & 4.3 & 21.3 & 1.0 \\[1.9pt]
		\multicolumn{2}{l}{Avg. Copper} & 422.9 & 9.9 & 4.7 & 6.4 & 36.3 & 0.8 \Bstrut\\\hline
	\end{tabular}
	\caption{Determined values of $R_{Pb}$ and $S_{{dust}}$ from Equation (\ref{eq:LinSystem}) for each sample and weighted averages. HDPE samples 7 and 8 were measured during a period of high noise in the UltraLo-1800 spectrometer and have been excluded from the analysis.\label{table:PostExp}}
\end{table}

\section{Discussion}

\subsection{Peak Activity}

With $R_{Pb}$ and $S_{{dust}}$ determined, one can predict the time at which maximum alpha activity occurs as a function of the exposure time:

\begin{equation}
t_{max}(t_{exp})=\frac{1}{\lambda_{Po}-\lambda_{Pb}}\ln\left[\frac{e^{(\lambda_{Po}t_{exp})}-1}{e^{(\lambda_{Pb}t_{exp})}-1}\right]
\label{eq:tmax}
\end{equation}

Using Equation (\ref{eq:tmax}) for $t$ in Equation (\ref{eq:ActTotal}) yields the maximum activity for any exposure time.  If a predetermined maximum activity is desired, one can solve this configuration of Equation (\ref{eq:ActTotal}) for $t_{exp}$ to determine the maximum exposure time.

\begin{figure}[h]
	\caption{Measurements of the HDPE samples plotted over the 95\% confidence interval of Equation (\ref{eq:ActBatemanClosed}) with $R_{Pb}$ and $S_{dust}$ taken from the weighted averages in Table \ref{table:PostExp}.}
	\includegraphics[width=\linewidth]{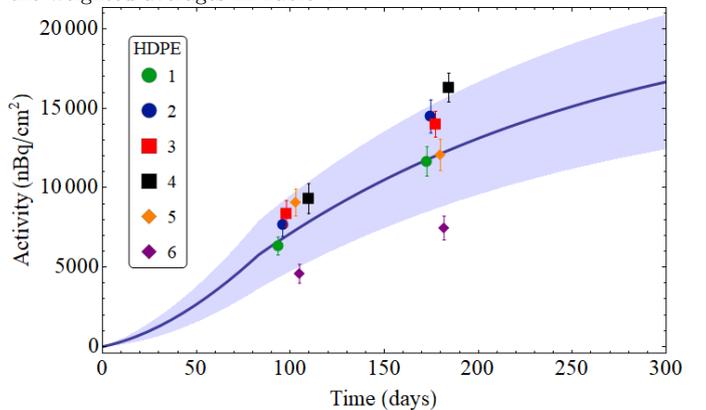}
	\label{im_AllPoints}
\end{figure}

\subsection{Contributions From Diffusion}

Radon can diffuse several millimeters into HDPE.  For
\makebox[\linewidth][s]{exposure times that are long relative to the half-life}
\makebox[\linewidth][s]{of $^{222}$Rn ($t_{1/2}=3.8$~d), $^{210}$Po activity will increase as}
$4.5\pm1.0$~pBq/cm$^2$/day/(Bq/m$^3$) \cite{Rau201265}.  From SRIM calculations, a 5.3~MeV alpha has a projected range in HDPE of 37~$\mu$m, leaving 4.5\% of the total diffusion-related activity within range to exit the bulk.  Of the alphas that manage to break the surface, 19\% will have an energy greater than the 2.0~MeV threshold used in this analysis.  Finally, the UltraLo-1800 only detects $\sim$50\% of all alphas (those emitted upwards).  For an 83~day exposure in the 135~Bq/m$^3$ SNOLAB environment, the total expected diffusion-related activity measured by the central electrode of the UltraLo-1800 spectrometer would be ${}0.013~\alpha$/day, corresponding to a determined activity of 0.4~nBq/cm$^2$, well below the initial activity of each panel (\textit{cf}. Table \ref{table:PreExp}). Moreover, after exposure at SNOLAB, all of the samples measured at greater than
100~$\alpha$/day (3~$\mu$Bq/cm$^2$);
thus diffusion-related activity is not expected to contribute to the overall determination of the $R_{Pb}$ and $S_{dust}$ parameters.

\subsection{Activity From Dust}


Previous studies in SNOLAB technical reports \cite{SNOLAB-STR-2007-003,SNOLAB-STR-2012-004} have looked at dust fallout and activity in the norite, shotcrete and concrete in the mine and lab, these materials being the likely components of dust in the lab. While SNOLAB advertises itself as a Class 2000 clean room, the measured particle-count levels from this study are consistent with Class 1000.  If an assumed typical Class 1000 dust fallout rate of 10 ng/hr/cm$^{2}$ is used with an average dust density of 2.5 g/cm$^3$, one would expect ${\sim}$96~nm/day of dust deposition, or an ${}$8~$\mu$m layer at the end of exposure in this study.

In determining the total expected activity per gram of dust, emanation efficiency losses are considered. Radon is likely to be flushed out by the argon gas flow in the XIA, so the alpha rate from $^{222}$Rn and its progeny is likely reduced by 20\% \cite{UNSCEAR}.  In the thorium series, $^{220}$Rn may decay while still in the XIA ($t_{1/2}=56$~s).  An assumed 75\% of these decays will plate out onto a nearby surface.  Taking the ratio of the sample size to the area of the spectrometer's tray, it is expected that the total alpha rate from $^{220}$Rn and its progeny will be reduced by 14\%. Based on the measured activity of norite, shotcrete and concrete at SNOLAB \cite{SNOLAB-STR-2007-003} and taking into account the activity reductions outlined above, the total U- and Th-chain alpha activity is $\sim$150\,Bq/kg, or $\sim$36 nBq/day/cm$^{2}$ from dust activity accumulation. 


A different activity measurement has been made for dirt from the vacuum cleaners in the clean room area of SNOLAB $\left(64.2 \textrm{ Bq/kg,  \cite{SNOLAB-STR-2012-004}}\right)$.  If this value is used for activity in dust, one would expect $\sim$15 nBq/day/cm$^{2}$ for dust activity accumulation.  Without knowing the filtration level of the vacuum filter or bag, this can be taken as a reasonable lower limit.  For polyethylene, our determined value of
\mbox{$24.9{\pm}4.3$\,nBq/day/cm$^{2}$} falls closer to the estimate based on rock activity and typical Class 1000 fallout rates.  The lower rate determined from the copper samples may be further evidence of surface roughness on the polyethylene samples, which could trap and hold dust particles better during shipping and handling as compared to the smoother surfaces of copper.

\subsection{Differences in HDPE and Copper}

The HDPE samples show a markedly lower plate-out rate for $^{210}$Pb and higher dust accumulation as compared to copper.  Surface roughness effects may explain both the higher dust-capture rate and the longer low-energy tails for the HDPE samples.  The higher plate-out rate for copper may indicate a higher bonding strength with radon progeny.  Additionally, the non-conducting surfaces that the samples laid on could have developed some static charge during initial handling which may increase the plate-out rate for copper.

One copper sample was cleaned with isopropyl alcohol and a third surface assay was performed with the UltraLo-1800 spectrometer.  We observed a modest 16\% reduction in activity relative to the surface-activity model derived from the first two assays.  Comparatively, an HDPE sample cleaned the same way showed a reduction of $\sim$90\% relative to the activity expected from its model.  The same copper sample was further cleaned with Radiacwash\texttrademark and deionized water, yielding a larger reduction of $\sim$60\% that is still fairly modest compared to the reduction observed for the HDPE sample.

\subsection{Locations Dependence}
There is not a strong case for suggesting a difference in plate-out rate for different locations, except possibly for the panels placed in Room 131 (\textit{cf}. Fig. \ref{im_map4}). While there are no doors or barriers between any of the sites, there are two large air handlers above Room 127 and Room 131.  The lower plate-out rate for sample 6 (\textit{cf}. Fig. \ref{im_AllPoints}) may be explained by a difference in the airflow rate immediately nearby the individual air handlers.  There does not seem to be a strong case for a difference in plate-out rate for samples placed at different heights or proximity to walls.

\subsection{Comparison to Jacobi Model Predictions}

A Jacobi model \cite{jacobi1972} prediction of plate-out height relies on a wide variety of parameters including the surface area and volume of the room, surface area of objects in the room, air ventilation rate, attachment rate of particles, detachment probability, and more.  Estimates for the ladder lab area of SNOLAB which see roughly 10 air changes per day give plate-out height estimates of 24--57~cm.  Our weighted averages for HDPE and copper fall on the low end of this estimate.

\section{Conclusions \& Outlook}

Through the exposure and subsequent assays of HDPE and copper samples, radon daughter plate-out and dust-activity accumulation values for HDPE and copper have been determined.  This allows for the creation of a predictive model of $^{210}$Pb contamination and activity from which exposure limits can be set to achieve specific background goals. 

Further studies could be conducted to determine how material type and surface roughness contribute to dust accumulation.  An additional study on cleaning techniques for these materials would also be useful, especially for the case of exposure beyond the desirable limit.  

\section{Acknowledgments}

The authors would like to extend their sincere gratitude to Kerry Loken for her support and assistance with this research. The authors would like to thank SNOLAB and its staff for support through underground space, logistical and technical services. SNOLAB operations are supported by the Canada Foundation for Innovation and the Province of Ontario Ministry of Research and Innovation, with underground access provided by Vale at the Creighton mine site. This work is supported in part by the United States Department of Energy, and by the National Science Foundation under grant number CAREER - NSF 1151869. Any opinions, findings, and conclusions or recommendations expressed in this material are those of the authors and do not necessarily reflect the views of the National Science Foundation. Fermilab is operated by the Fermi Research Alliance, LLC under Contract No. De-AC02-07CH11359. Pacific Northwest National Laboratory is operated by Battelle for the United States Department of Energy under Contract No. DE-AC05-76RL01830.


\bibliography{myreferences}

\end{document}